\documentclass[conference]{IEEEtran} 

\IEEEoverridecommandlockouts

\usepackage{graphics} % for pdf, bitmapped graphics files
\usepackage{subcaption}
\usepackage{placeins}
\usepackage{epsfig} % for postscript graphics files
\usepackage{amsmath} % assumes amsmath package installed
\usepackage{amssymb}  % assumes amsmath package installed

\usepackage{url}
\usepackage{hyperref}
\usepackage[ruled, vlined, linesnumbered]{algorithm2e}
\usepackage{verbatim} 
\usepackage{soul, color}
\usepackage{lmodern}
\usepackage{fancyhdr}
\usepackage[utf8]{inputenc}
\usepackage{fourier} 
\usepackage{array}
\usepackage{makecell}

\SetNlSty{large}{}{:}

\makeatletter

\newcommand{\Rom}[1]{\expandafter\@slowromancap\romannumeral #1@}
\makeatother

\title{\LARGE \bf
A Reversible Continuous-Variable Photonic Memory Architecture Based on Displacement-Evolved Coherent-State Dynamics
}

\author{
\IEEEauthorblockN{Sanjit Krishna}
\IEEEauthorblockA{
Department of Physics \\
National Institute of Technology
Surat \\
Surat,Gujarat India \\
i24ph006@phy.svnit.ac.in
}
}

\begin{document}

\thispagestyle{plain}
\pagestyle{plain}

\maketitle

%%%%%%%%%%%%%%%%%%%%%%%%%%%%%%%%%%%%%%%%%%%%%%%%%%%%%%%%%%%%%%%%%%%%%%%%%%%%%%%%
\begin{abstract}

Modern information systems increasingly require memory architectures capable of storing not only static data but also the evolution of information over time. While continuous-variable photonic platforms have been extensively studied for quantum communication and computation, their potential as dynamically evolving memory systems remains largely unexplored. In this work, we introduce a metadata-assisted continuous-variable photonic memory architecture in which information is represented by latent vectors encoded across multimode coherent states and updated through reversible displacement operations \cite{sanjit2025}. The framework combines multimode storage with metadata-based indexing, enabling navigation and reconstruction of historical memory states through rollback retrieval.
\vspace{0.1cm}

The proposed architecture is analyzed using an eight-qumode coherent-state model incorporating optical loss, amplifier-assisted compensation, and stochastic noise. Numerical simulations demonstrate stable memory evolution under ideal conditions and quantify retrieval degradation under realistic noise environments. Retrieval fidelity remains high in the low-noise regime and exhibits predictable decay with accumulated noise, while rollback reconstruction preserves substantial information even at significant retrieval depths. Analytical estimates of memory lifetime and Rollback Distinguishability Capacity further characterize the scaling behavior of the architecture.
\vspace{0.1cm}

Unlike conventional approaches that focus on preserving a single quantum state, the proposed framework treats memory as a continuously evolving trajectory in continuous-variable phase space.\cite{braunstein2005} This perspective naturally extends beyond image storage to temporal data streams, machine-learning latent representations, scientific simulations, and other forms of dynamically changing information. Although presented as a theoretical proof-of-concept, the results suggest that continuous-variable photonic systems may provide a promising foundation for future metadata-aware memory architectures capable of storing, tracking, and reconstructing the history of evolving information.
\vspace{0.1cm}

\end{abstract}

%%%%%%%%%%%%%%%%%%%%%%%%%%%%%%%%%%%%%%%%%%%%%%%%%%%%%%%%%%%%%%%%%%%%%%%%%%%%%%%%
\section{INTRODUCTION}
Photonic platforms promise high-bandwidth,low noise information processing, but they lack inherent memory to buffer signals. An optical quantum memory, even a simple delay line or fiber loop, is often needed to synchronize and reorder photons. Optical signals can be delayed by passive loops or resonators, and switchable delay lines have been proposed to synchronize probabilistic sources. But existing memories face severe practical limits: atomic-ensemble memories demand narrowband control fields and often cryogenic setups, while fixed fiber delays only provide a preset latency and suffer unavoidable loss and dispersion. Furthermore, continuous-variable proposals typically assume ideal lossless loops and negligible control overhead. In particular, addressing many time-multiplexed modes is hard in practice even gigahertz rate electro-optic modulators can only target a few modes per round-trip, making large-$M$ (modes) schemes unscalable.
\vspace{0.1cm}

To overcome these gaps, we propose a reversible multi-mode coherent-state memory that stores only a fixed small set of $M$ optical modes in a circulating loop, rather than one mode per pixel or per pulse.  In this architecture, an input frame is first compressed into $M\ll N$ latent features via a classical encoder. These features modulate $M$ coherent-state pulses that circulate continuously in a multi mode optical loop. At each round-trip, an electro-optic modulator applies a small displacement $D(\Delta\alpha_i)$ to each circulating pulse $|\alpha_i\rangle$, evolving the memory under new inputs. Inside the loop we include an optical parametric amplifier (OPA) and a phase-locked local oscillator to compensate round-trip loss ($\eta<1$) and phase drift. A weak beam-splitter taps a tiny fraction of each pulse for occasional homodyne readout. We do not continuously collapse the memory instead, a classical controller here referred to as a digital-twin uses the sparse homodyne samples to calibrate a classical simulation of the loop. The controller stores the intended displacement updates $\{\Delta\alpha_i^{(n)}\}$ and measured metadata (e.g. entropy of quadrature noise) in memory. Later, to recover an earlier frame we apply the inverse displacements $D(-\Delta\alpha_i)$ in reverse order, approximately undoing the evolution.
\vspace{0.1cm}

Theoretically, our model predicts the stored amplitudes to evolve as 
\[
\tilde{\alpha}_{n+1} = \sqrt{\eta}\,\tilde{\alpha}_n + \Delta\alpha_n +\epsilon_i^{(n)},
\] 
where $\eta$ is the loop round-trip gain (including OPA correction) and $\epsilon_i^{(n)},$ is accumulated Gaussian noise. The ideal (lossless) trajectory would be $\alpha_n^{\rm ideal}=\alpha_0+\sum_{k<n}\Delta\alpha_k$.  We define the error $\delta_n=\alpha_n^{\rm ideal}-\tilde{\alpha}_n$ so that the fidelity between ideal and actual coherent states is 
$$
F_n \;=\;|\langle \alpha_n^{\rm ideal}|\tilde{\alpha}_n\rangle|^2
\;=\;\exp\big(-|\delta_n|^2\big)\,. 
$$ 
From this one can compute a memory lifetime as the number of steps until $F_n$ falls below a threshold (or entropy accumulates beyond a limit).  These analytical results allow comparison between an idealized model (no loss, perfect control) and a realistic scenario (finite $\eta<1$, amplifier noise, discrete feedback, etc.)
\vspace{0.1cm}

Existing continuous-variable and photonic quantum memory architectures primarily focus on preserving the fidelity of the most recently stored quantum state, with comparatively little attention given to reconstructing historical memory evolution. In contrast, the proposed framework introduces a metadata-assisted rollback mechanism that enables reversible retrieval of previously stored coherent-state memory states without requiring multiple physical copies of the stored information. Furthermore, we develop a theoretical rollback retrieval fidelity that quantifies reconstruction accuracy under realistic physical perturbations and digital-twin estimation uncertainty, and introduce a rollback capacity based on rollback distinguishability to characterize the maximum number of mutually retrievable historical memory states.
\vspace{0.1cm}

The digital twin is not intended to replace the physical photonic memory. Instead, it serves as a monitoring, indexing, and retrieval layer, while the optical loop performs the real-time coherent-state evolution at photonic bandwidth. In the present work, coherent states permit efficient digital tracking; however, the same architecture naturally extends to non-classical continuous-variable states, where exact classical simulation is no longer efficient and the physical photonic memory provides the primary computational advantage. Although the present implementation employs coherent states, future extensions to squeezed and entangled continuous-variable states are expected to provide genuine quantum advantages beyond efficient classical simulation.

\vspace{0.1cm}

\section{Encoding and Storage Framework}

\subsection{Background: Coherent States and Displacement Operators}

The proposed memory architecture is based on continuous-variable (CV) photonic quantum systems. In the CV formalism, each optical mode is modeled as a quantum harmonic oscillator described by the bosonic annihilation and creation operators $\hat{a}$ and $\hat{a}^{\dagger}$, satisfying

\begin{equation}
[\hat{a},\hat{a}^{\dagger}] = 1.
\end{equation}

The Fock states $|n\rangle$ satisfy

\begin{equation}
\hat{a}^{\dagger}\hat{a}|n\rangle = n|n\rangle .
\end{equation}

A coherent state $|\alpha\rangle$ is defined as the eigenstate of the annihilation operator,

\begin{equation}
\hat{a}|\alpha\rangle = \alpha|\alpha\rangle .
\end{equation}

The coherent state can be generated from the vacuum state through the displacement operator \cite{gerry2005}

\begin{equation}
D(\alpha)
=
\exp \left(
\alpha \hat{a}^{\dagger}
-
\alpha^{*}\hat{a}
\right),
\end{equation}

such that

\begin{equation}
|\alpha\rangle
=
D(\alpha)|0\rangle .
\end{equation}

A key property of the displacement operator is

\begin{equation}
D(\beta)|\alpha\rangle
=
|\alpha+\beta\rangle ,
\end{equation}

which enables controlled evolution of coherent-state amplitudes.

In this work, each coherent optical mode is referred to as a qumode. A multimode coherent-state system containing $M$ qumodes is represented by

\begin{equation}
|\Psi\rangle
=
\bigotimes_{i=1}^{M}
|\alpha_i\rangle .
\end{equation}

\vspace{0.1cm}
Rather than assigning one qumode to every pixel, a fixed number of latent qumodes ($M=8,16,$ or $32$) is used to encode the dominant visual features of the image.

\subsection{Feature Encoding}

The proposed memory architecture does not store individual image pixels directly because that would require the number of qumodes that scales up with the image size. So, each image frame is transformed into a compact latent representation that captures the dominant visual characteristics of the image. This significantly reduces the number of required qumodes while preserving the information necessary for subsequent retrieval and rollback reconstruction.

Let

\begin{equation}
I^{(n)}(x,y)
\end{equation}

denote an image frame acquired at time step $n$, where $x$ and $y$ represent the spatial coordinates of the image.
\vspace{0.1cm}

To ensure deterministic processing, the image is traversed using any conventional feature-encoding procedure.The acquired image information is subsequently transformed into a latent feature vector

\begin{equation}
\mathbf{c}^{(n)}
=
\left[
c_1^{(n)},
c_2^{(n)},
\ldots,
c_M^{(n)}
\right],
\end{equation}

where $M$ denotes the number of latent feature channels.
\vspace{0.1cm}

The extracted coefficients correspond to dominant image characteristics such as brightness, contrast, edge information, texture content, temporal variation, entropy, and residual image information. For example, the mean image intensity is defined as

\begin{equation}
c_1^{(n)}
=
\frac{1}{WH}
\sum_{x=1}^{W}
\sum_{y=1}^{H}
I^{(n)}(x,y),
\end{equation}

where $W$ and $H$ denote the width and the height of image, while the image contrast is given by

\begin{equation}
c_2^{(n)}
=
\sqrt{
\frac{1}{WH}
\sum_{x=1}^{W}
\sum_{y=1}^{H}
\left(
I^{(n)}(x,y)-c_1^{(n)}
\right)^2
}.
\end{equation}

The information content of the image may be quantified through the Shannon entropy

\begin{equation}
c_{\mathrm{ent}}^{(n)}
=
-
\sum_{k}
p_k^{(n)}
\log p_k^{(n)},
\end{equation}

where $p_k^{(n)}$ denotes the  probability of obtaining the $k$th intensity level.
\vspace{0.1cm}

Rather than assigning one qumode to every image pixel, this architecture assigns one qumode to each latent feature channel. This allows the number of required qumodes remain fixed and independent of image resolution. The implementations may employ $M=8$, $M=16$, or $M=32$ latent qumodes depending on the desired balance between storage efficiency and reconstruction fidelity.

Each latent coefficient is subsequently mapped onto a coherent-state amplitude according to

\begin{equation}
c_i^{(n)}
\rightarrow
\alpha_i^{(n)},
\qquad
i=1,2,\ldots,M.
\end{equation}

The multi-mode coherent-state representation of the image frame is therefore

\begin{equation}
|\Psi^{(n)}\rangle
=
\bigotimes_{i=1}^{M}
|\alpha_i^{(n)}\rangle.
\end{equation}

\subsection{Displacement-Evolution Storage Architecture}

After feature extraction and coherent-state encoding, the resulting multimode state is injected into a time-multiplexed optical loop memory. Rather than generating a new set of qumodes for every incoming image frame, the proposed architecture continuously updates a fixed set of qumodes through controlled displacement operations.

Consider a multimode coherent-state representation

\begin{equation}
|\Psi^{(n)}\rangle
=
\bigotimes_{i=1}^{M}
|\alpha_i^{(n)}\rangle,
\end{equation}

where $M$ denotes the number of latent qumodes and $n$ denotes the frame index.

The qumodes are stored as a train of temporally separated optical pulses. If $\tau$ denotes the temporal separation between adjacent pulses, then the arrival time of the $i$th pulse is

\begin{equation}
t_i = i\tau,
\qquad
i = 1,2,\ldots,M.
\end{equation}

The optical loop is designed such that one complete round trip corresponds to

\begin{equation}
T_{\mathrm{loop}}
=
M\tau.
\end{equation}

Hence, each pulse returns to the same temporal slot after every loop circulation, enabling synchronized displacement updates through an electro-optic modulator controlled by classical metadata.\cite{ghatak1998}
\vspace{0.1cm}

For two consecutive image frames, let the coherent-state amplitudes be represented by $\alpha_i^{(n)}$ and $\alpha_i^{(n+1)}$, respectively. The required displacement increment is defined as

\begin{equation}
\Delta\alpha_i^{(n)}
=
\alpha_i^{(n+1)}
-
\alpha_i^{(n)}.
\end{equation}

When the $i$th pulse reaches the electro-optic modulator at its assigned temporal slot, a displacement operation

\begin{equation}
D\!\left(\Delta\alpha_i^{(n)}\right)
\end{equation}

is applied.

Using the coherent-state displacement relation

\begin{equation}
D(\beta)|\alpha\rangle
=
|\alpha+\beta\rangle,
\end{equation}

the updated state becomes

\begin{equation}
|\alpha_i^{(n+1)}\rangle
=
D\!\left(\Delta\alpha_i^{(n)}\right)
|\alpha_i^{(n)}\rangle.
\end{equation}

Substituting Eq.~(20) into Eq.~(21), one obtains

\begin{equation}
|\alpha_i^{(n+1)}\rangle
=
|\alpha_i^{(n)}
+
\Delta\alpha_i^{(n)}
\rangle.
\end{equation}

So, the memory evolves through incremental displacement updates rather than complete state replacement. This enables temporal information to be stored efficiently while maintaining a fixed number of physical qumodes.
\vspace{0.1cm}

In a realistic optical loop, attenuation and environmental noise lead to deviations from ideal evolution as discussed before. Let $\eta$ denote the loop transmission coefficient, where $0<\eta<1$. The practical evolution of the coherent-state amplitude may then be modeled as

\begin{equation}
\tilde{\alpha}_i^{(n+1)}
=
\sqrt{\eta}\,
\tilde{\alpha}_i^{(n)}
+
\Delta\alpha_i^{(n)}
+
\epsilon_i^{(n)},
\end{equation}

where $\epsilon_i^{(n)}$ represents accumulated noise introduced by optical loss, phase fluctuations, and imperfect modulation. This will also be discussed in detail in the next section.
\vspace{0.1cm}

The displacement history $\Delta\alpha_i^{(n)}$ is stored classically for every qumode and every time step. This stored metadata enables subsequent rollback operations through inverse displacements, allowing approximate reconstruction of previously stored states without requiring the preservation of all intermediate optical states.

\subsection{Metadata Management and Retrieval Indexing}

The optical loop stores only the current evolved coherent-state configuration. Hence, additional information required for retrieval and rollback operations is maintained in a classical metadata database.
\vspace{0.1cm}

The metadata database is assumed to be updated through sparse homodyne monitoring of the multimode photonic memory. Rather than performing full state tomography after every update, periodic quadrature measurements provide coarse estimates of the evolving coherent-state amplitudes. These estimates, together with displacement histories and frame indices, form a digital twin of the memory evolution and enable metadata-assisted rollback retrieval. \cite{silverstone2016}

For each frame \(n\), the stored metadata is represented by

\begin{equation}
\mathcal{M}^{(n)}
=
\left\{
\Delta\alpha_i^{(n)},
S^{(n)},
F^{(n)},
t_n
\right\},
\end{equation}

where \(\Delta\alpha_i^{(n)}\) denotes the displacement history of the \(i\)th qumode, \(S^{(n)}\) is the image entropy, \(F^{(n)}\) is the estimated storage fidelity, and \(t_n\) represents the frame timestamp.
\vspace{0.1cm}

The entropy tag is computed classically from the image intensity histogram according to

\begin{equation}
S^{(n)}
=
-
\sum_k
p_k^{(n)}
\log p_k^{(n)},
\end{equation}

where \(p_k^{(n)}\) denotes the probability associated with the \(k\)th intensity level. This quantity provides a compact measure of image complexity and information content, thereby enabling efficient indexing and retrieval of stored frames.
\vspace{0.1cm}

During retrieval, the metadata database is first searched using the desired frame index, timestamp, or entropy tag. The corresponding displacement history is then loaded and used to generate a sequence of inverse displacement operations

\begin{equation}
D^{-1}
\!\left(
\Delta\alpha_i^{(n)}
\right)
=
D
\!\left(
-\Delta\alpha_i^{(n)}
\right),
\end{equation}

which enables rollback of the stored coherent-state trajectory. Consequently, entropy serves as a retrieval descriptor, while the displacement history provides the information required for approximate reconstruction of previously stored states.

\vspace{0.1cm}

\section{Noise Modeling and Rollback Reconstruction}

\subsection{Practical Optical Loop Considerations}

The storage model discussed in the previous section assumes ideal coherent state evolution. In a practical implementation, the circulating optical pulses experience attenuation, phase drift, and modulation imperfections during repeated propagation through the optical loop. These effects must be considered when analyzing long-term storage and rollback performance.\cite{optec}
\vspace{0.1cm}

The dominant source arises from optical attenuation in the fiber delay line, beam splitters, couplers, and other passive photonic components. For a loop transmission coefficient $\eta$, where $0 < \eta < 1$, the coherent-state amplitude evolves as

\begin{equation}
|\alpha\rangle
\rightarrow
|\sqrt{\eta}\,\alpha\rangle.
\end{equation}

Also, repeated loop circulations gradually reduce the stored optical amplitude and may limit the achievable storage duration.

To mitigate this effect, an optical parametric amplifier (OPA) may be incorporated within the loop. Let $G$ denote the amplifier gain. Ideally, the OPA transforms the coherent state according to

\begin{equation}
|\alpha\rangle
\rightarrow
|\sqrt{G}\,\alpha\rangle.
\end{equation}

Under suitable operating conditions, the gain may be selected such that

\begin{equation}
G \approx \frac{1}{\eta},
\end{equation}

so it partially compensates the attenuation accumulated during each round trip. Nevertheless, practical amplifiers introduce excess noise and cannot perfectly restore the original state.

\vspace{0.1cm}

Apart from attenuation, coherent-state storage is sensitive to phase fluctuations. Since information is encoded in both the amplitude and phase of the coherent-state parameter $\alpha$, small phase deviations accumulated during propagation can lead to reconstruction errors during retrieval. Therefore, active phase stabilization techniques and phase-locked reference oscillators are assumed to maintain stable operation of the optical loop.\cite{caves1982}

\vspace{0.1cm}

The combined influence of attenuation, amplifier noise, phase fluctuations, and imperfect displacement operations is represented by an effective noise contribution $\epsilon_i^{(n)}$. This perturbation term represents the cumulative effect of multiple statistically independent physical noise sources encountered during each memory update, including optical loss, phase fluctuations, modulator imperfections, detector uncertainty, and quantum-limited amplifier noise. Accordingly,

\begin{equation}
\epsilon_i^{(n)}
=
\epsilon_{\mathrm{loss}}^{(n)}
+
\epsilon_{\mathrm{phase}}^{(n)}
+
\epsilon_{\mathrm{OPA}}^{(n)}
+
\epsilon_{\mathrm{mod}}^{(n)}
+
\epsilon_{\mathrm{det}}^{(n)}
+\cdots .
\end{equation}

For the optical parametric amplifier (OPA), the amplifier contribution is fundamentally constrained by the quantum limit established by Caves' theorem, which requires every phase-insensitive amplifier to introduce a minimum vacuum-noise contribution. Consequently, $\epsilon_{\mathrm{OPA}}^{(n)}$ incorporates both technical amplifier fluctuations and the irreducible quantum noise associated with amplification.
\vspace{0.1cm}

Since the overall perturbation is the sum of several independent stochastic processes, the Central Limit Theorem justifies modelling the aggregate perturbation as an additive zero-mean Gaussian random variable,

\begin{equation}
\epsilon_i^{(n)}
\sim
\mathcal{N}
\left(
0,\sigma_{\epsilon}^{2}
\right),
\end{equation}

where the effective variance

\begin{equation}
\sigma_{\epsilon}^{2}
=
\sigma_{\mathrm{loss}}^{2}
+
\sigma_{\mathrm{phase}}^{2}
+
\sigma_{\mathrm{OPA}}^{2}
+
\sigma_{\mathrm{mod}}^{2}
+
\sigma_{\mathrm{det}}^{2}
+\cdots
\end{equation}

captures the combined contribution of all physical noise mechanisms throughout the memory evolution.
\vspace{0.1cm}

The practical evolution of the $i^{\text{th}}$ qumode may therefore be expressed as

\begin{equation}
\tilde{\alpha}_i^{(n+1)}
=
\sqrt{\eta{G}}\,
\tilde{\alpha}_i^{(n)}
+
\Delta\alpha_i^{(n)}
+
\epsilon_i^{(n)},
\end{equation}

where $\tilde{\alpha}_i^{(n)}$ denotes the actual stored amplitude after accounting for physical imperfections. This expression forms the basis for the rollback and reconstruction analysis presented in the following subsections.

\subsection{Digital-Twin Assisted Rollback Retrieval}

At any instant, the photonic memory contains only the most recently evolved coherent-state configuration. To preserve access to the complete temporal evolution without disturbing the optical memory, the proposed architecture maintains a digital twin that continuously tracks the evolution of the multimode coherent-state amplitudes using sparse homodyne measurements and the recorded displacement history.
\vspace{0.1cm}

For every stored frame, the metadata database maintains the estimated coherent-state amplitudes together with the associated displacement operations, entropy descriptor, fidelity estimate, and temporal index. The metadata associated with the $n^{\mathrm{th}}$ frame may therefore be written as

\begin{equation}
\mathcal{M}^{(n)}
=
\left\{
\tilde{\alpha}_i^{(n)},
\Delta\alpha_i^{(n)},
S^{(n)},
F^{(n)},
t_n
\right\},
\end{equation}

where $\tilde{\alpha}_i^{(n)}$ denotes the estimated coherent-state amplitudes stored by the digital twin.
\vspace{0.1cm}

The digital twin evolves simultaneously with the optical memory according to

\begin{equation}
\tilde{\alpha}_i^{(n+1)}
=
\sqrt{\eta G}\,
\tilde{\alpha}_i^{(n)}
+
\Delta\alpha_i^{(n)}
+
\epsilon_i^{(n)},
\end{equation}

where $\epsilon_i^{(n)}$ represents the accumulated perturbation introduced during the memory evolution, including the combined effects of optical loss, amplifier fluctuations, phase noise, modulation imperfections, and other stochastic disturbances. Throughout this work, these cumulative perturbations are modelled as additive zero-mean Gaussian noise.Since coherent states remain coherent under linear loss, amplification, and displacement operations, the digital twin is required to track only the complex coherent-state amplitudes rather than the complete quantum state, thereby significantly reducing the computational resources required for storage and retrieval.
\vspace{0.1cm}

Suppose a user requests a frame indexed by $m$. The corresponding metadata entry is first identified using the stored frame index, timestamp, or entropy descriptor. The entropy

\begin{equation}
S^{(m)}
=
-
\sum_k
p_k^{(m)}
\log p_k^{(m)}
\end{equation}

acts as a compact descriptor of image complexity and enables efficient indexing of visually similar frames.
\vspace{0.1cm}
Once the desired frame has been identified, the stored coherent-state amplitudes contained in the digital twin are directly mapped back to the latent feature coefficients and decoded to reconstruct the corresponding image frame. Consequently, the complete image or video sequence may be retrieved entirely from the digital twin without modifying the physical photonic memory, which continues to preserve only the most recent coherent-state configuration.
\vspace{0.1cm}

The displacement history recorded within the metadata additionally provides a reversible description of the optical evolution. Accordingly, the corresponding inverse displacement operator is

\begin{equation}
D^{-1}
\left(
\Delta\alpha_i^{(n)}
\right)
=
D
\left(
-\Delta\alpha_i^{(n)}
\right).
\end{equation}

Although image and video reconstruction are performed directly from the digital twin, the recorded displacement history also enables rollback of the \emph{physical} photonic memory whenever retrieval of an actual historical coherent state is required. In this case, successive inverse displacement operations may be applied to the physical memory to recover earlier optical states while preserving the underlying coherent-state dynamics. The analytical description of the resulting rollback state, denoted by $\alpha_{\mathrm{RB}}$, together with the corresponding rollback retrieval fidelity $F_{\mathrm{RB}}$, is developed in the subsequent sections.
\vspace{0.1cm}

\subsection{Reconstruction Error and Retrieval Fidelity}

The rollback reconstruction process is subject to imperfections arising from optical attenuation, amplifier noise, phase fluctuations, and inaccuracies in displacement operations. Consequently, the reconstructed coherent-state amplitude differ from the ideal stored amplitude.

For the $i$th qumode, the reconstruction error is defined as

\begin{equation}
\delta_i^{(m)}
=
\alpha_i^{(m)}
-
\hat{\alpha}_i^{(m)},
\end{equation}

where $\alpha_i^{(m)}$ denotes the ideal coherent-state amplitude corresponding to frame $m$, and $\hat{\alpha}_i^{(m)}$ represents the amplitude reconstructed by the digital twin.

The retrieval accuracy is quantified using the coherent-state fidelity. For a single qumode, the fidelity is given by

\begin{equation}
F_i^{(m)}
=
\left|
\left<
\alpha_i^{(m)}
\middle|
\hat{\alpha}_i^{(m)}
\right>
\right|^2.
\end{equation}

Using the coherent-state overlap relation derived previously, the fidelity may be expressed as

\begin{equation}
F_i^{(m)}
=
\exp
\left(
-
\left|
\delta_i^{(m)}
\right|^2
\right).
\end{equation}

For an $M$-qumode memory, the overall retrieval fidelity is

\begin{equation}
F_{\mathrm{tot}}^{(m)}
=
\prod_{i=1}^{M}
F_i^{(m)}.
\end{equation}

As the storage duration and rollback depth increase, accumulated imperfections generally lead to larger reconstruction errors and a corresponding reduction in retrieval fidelity. The behaviour of these quantities under realistic operating conditions is investigated through numerical simulations in the following section.

\section{Numerical Simulations and Analysis}
\subsection{Evolution of Stored Coherent States}

To investigate the storage dynamics of the proposed continuous-variable photonic memory, numerical simulations were performed for an eight-qumode architecture undergoing successive displacement updates. Each qumode stores a latent feature coefficient encoded as the amplitude of a coherent state. As new information arrives, displacement operations modify the stored amplitudes, producing an evolving trajectory in the latent-state space.\cite{weedbrook2012}

Figure $1$ compares the evolution of the coherent state amplitudes under ideal and realistic operating conditions. In the ideal case shown in Fig.$1a$, loss and noise are neglected and the effective gain-loss product satisfies $G\eta = 1$. So, the coherent amplitudes evolve exclusively through the displacement updates,

\begin{equation}
\alpha_{n+1}
=
\alpha_n+\Delta\alpha_n,
\end{equation}

where $\Delta\alpha_n$ represents the latent feature update associated with the incoming image frame. The resulting trajectories exhibit smooth evolution across all eight qumodes, demonstrating the accumulation of information within the memory architecture.
\vspace{0.1cm}

In practice, optical propagation losses, modulator imperfections, and amplifier noise perturb the stored amplitudes. To model these effects, the memory evolution is described by

\begin{equation}
\tilde{\alpha}_{n+1}
=
\sqrt{G\eta}\,\tilde{\alpha}_n
+
\Delta\alpha_n
+
\epsilon_i^{(n)},
\label{eq:noisy_evolution_sim}
\end{equation}

where $\eta$ denotes the transmission efficiency of the optical loop, $G$ is the amplifier gain, and $\epsilon_i^{(n)},$ is a complex Gaussian noise term representing amplifier and control noise. Fig.$1b$ shows the corresponding evolution under these realistic conditions.

As compared to the ideal trajectories, the noisy evolution exhibits fluctuations and gradual deviations arising from the accumulation of stochastic perturbations. However, the overall latent-state structure remains preserved throughout the storage process. This behavior indicates that amplifier-assisted compensation successfully mitigates excessive attenuation while maintaining stable memory operation over extended storage durations.
\vspace{0.2cm}

The comparison between this two cases highlights the robustness of the displacement-based storage mechanism. While loss and noise perturb the coherent amplitudes, the encoded latent information remains recoverable, motivating the fidelity and rollback analyses presented in the following subsections.
\vspace{0.2cm}

From an information-storage perspective, the trajectories indicate that the proposed multimode architecture is capable of continuously incorporating new latent updates while retaining previously stored information.
\begin{figure}[!htbp]
\centering

\begin{subfigure}{0.8\linewidth}
\centering
\includegraphics[width=\linewidth]{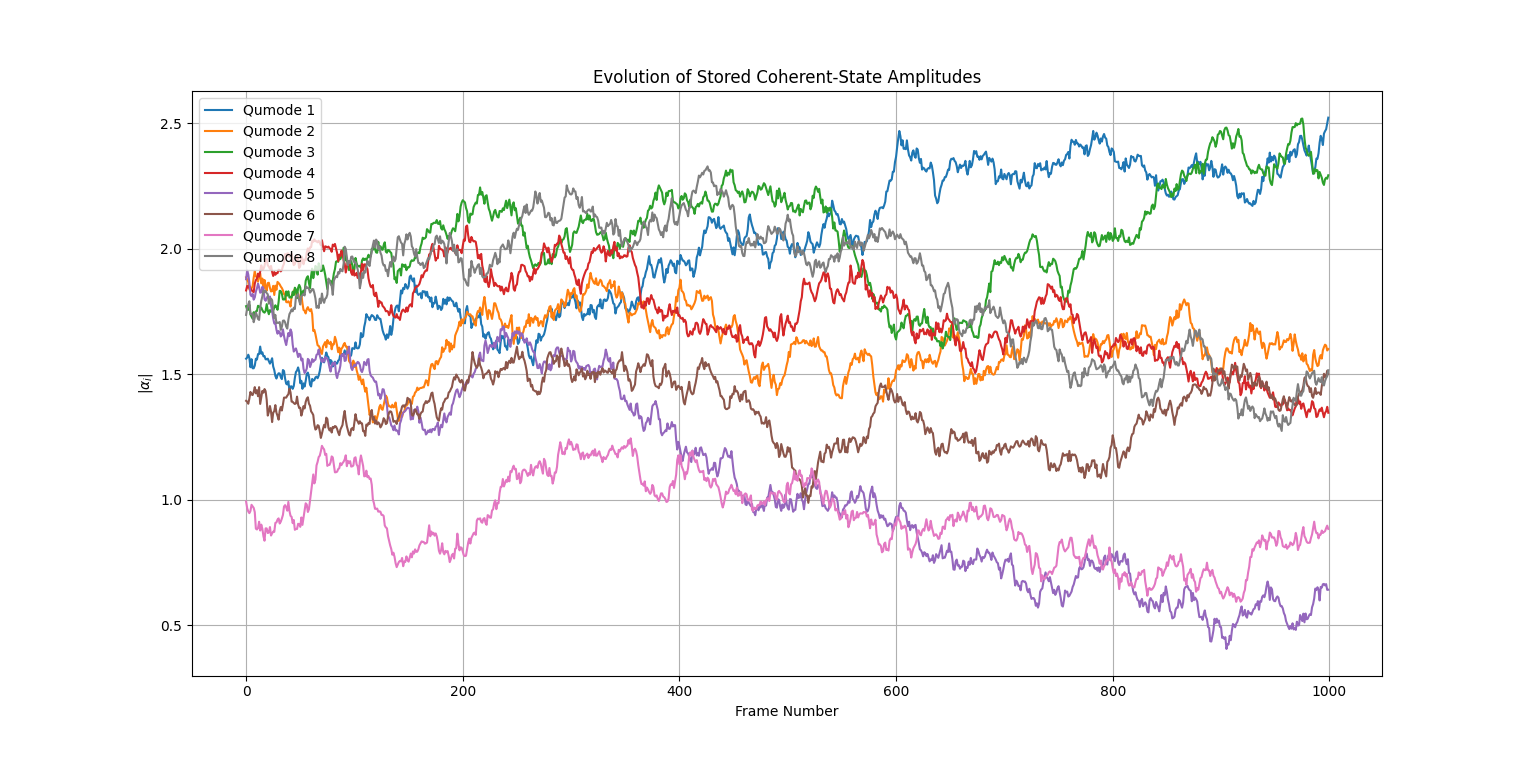}
\caption{Ideal evolution ($G\eta=1$).}
\end{subfigure}
\hfill
\begin{subfigure}{0.8\linewidth}
\centering
\includegraphics[width=\linewidth]{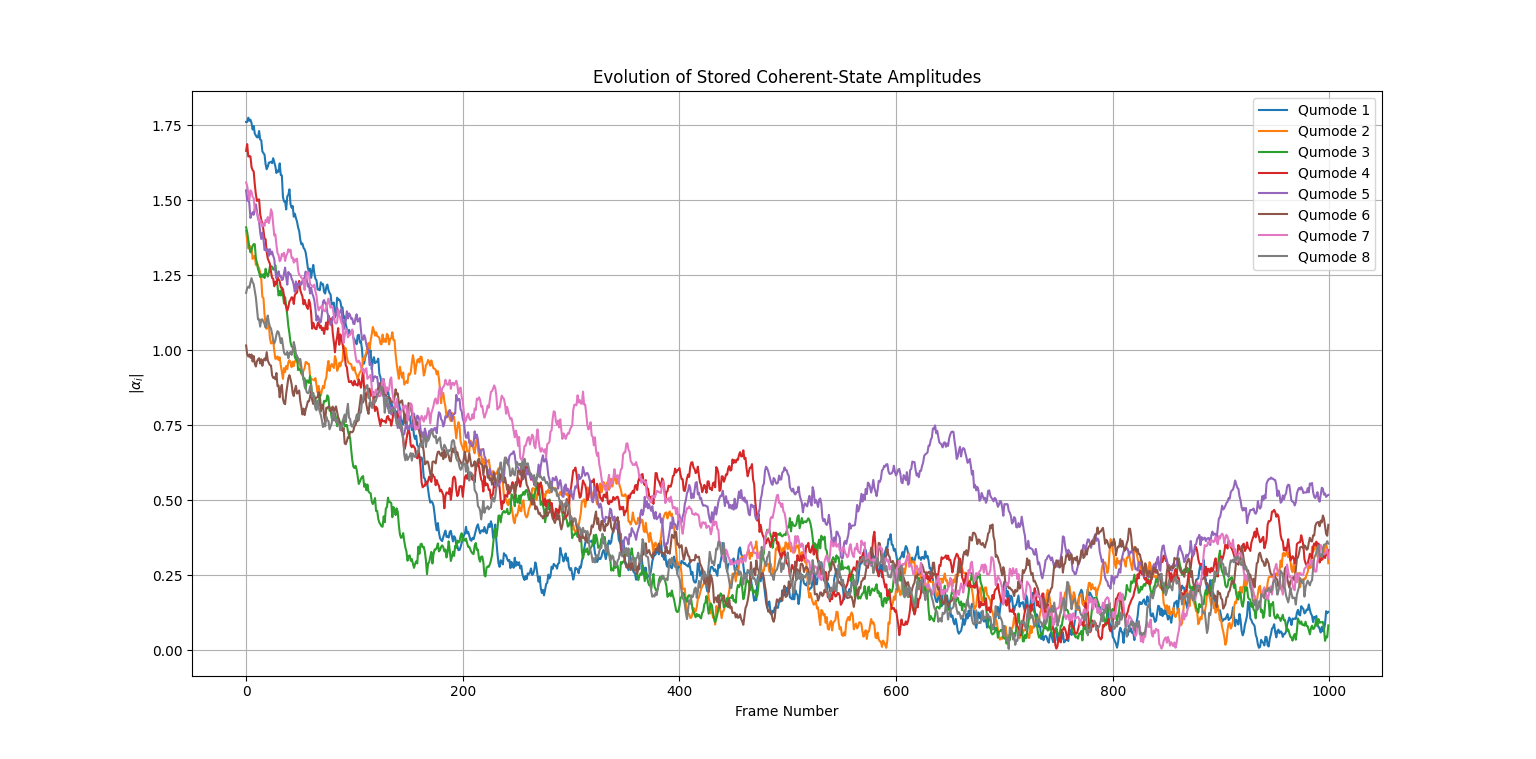}
\caption{Noisy evolution ($G\eta<1$).}
\end{subfigure}

\caption{
Comparison of coherent-state evolution in the proposed eight-qumode photonic memory architecture. (a) Ideal evolution in the absence of loss and noise, where the coherent amplitudes evolve solely through displacement updates. (b) Evolution under realistic operating conditions including optical loss, amplifier compensation, and noise. Although fluctuations accumulate over time, the latent-state structure remains preserved, demonstrating the robustness of the storage mechanism.
}
\label{fig:evolution_comparison}

\end{figure}

\FloatBarrier
\subsection{Retrieval Fidelity Under Different Noise Levels}

To quantify the impact of noise on memory performance, the retrieval fidelity was evaluated under different levels of stochastic perturbation. The simulations were performed for an eight-qumode memory over 1000 storage frames, and the results were averaged over 50 independent realizations to reduce statistical fluctuations.
\vspace{0.1cm}

The retrieval fidelity at frame $n$ is defined as the overlap between the ideal coherent-state representation and the corresponding noisy stored state,

\begin{equation}
F_n
=
\left|
\left\langle
\alpha_n^{\mathrm{ideal}}
\middle|
\tilde{\alpha}_n
\right\rangle
\right|^2,
\end{equation}

where $\alpha_n^{\mathrm{ideal}}$ denotes the ideal lossless coherent-state amplitude and $\tilde{\alpha}_n$ represents the corresponding amplitude obtained under realistic operating conditions. For coherent states, this overlap can be expressed as

\begin{equation}
F_n
=
\exp
\left(
-
\left|
\alpha_n^{\mathrm{ideal}}
-
\tilde{\alpha}_n
\right|^2
\right).
\end{equation}

The retrieval fidelity presented in this subsection characterizes the cumulative degradation of the stored coherent-state representation during the forward evolution of the photonic memory. It therefore quantifies the accuracy with which the most recently stored frame can be recovered directly from the memory. In contrast, the rollback retrieval fidelity introduced in the following subsection evaluates the accuracy of reconstructing a historical memory state through successive inverse displacement operations. Consequently, retrieval fidelity measures the robustness of the storage process itself, whereas rollback retrieval fidelity quantifies the stability of the proposed rollback mechanism for accessing earlier memory states. Together, these two metrics provide complementary measures of the performance of the proposed photonic memory framework.
\vspace{0.1cm}

\begin{figure}[!t]
\centering
\includegraphics[width=\columnwidth]{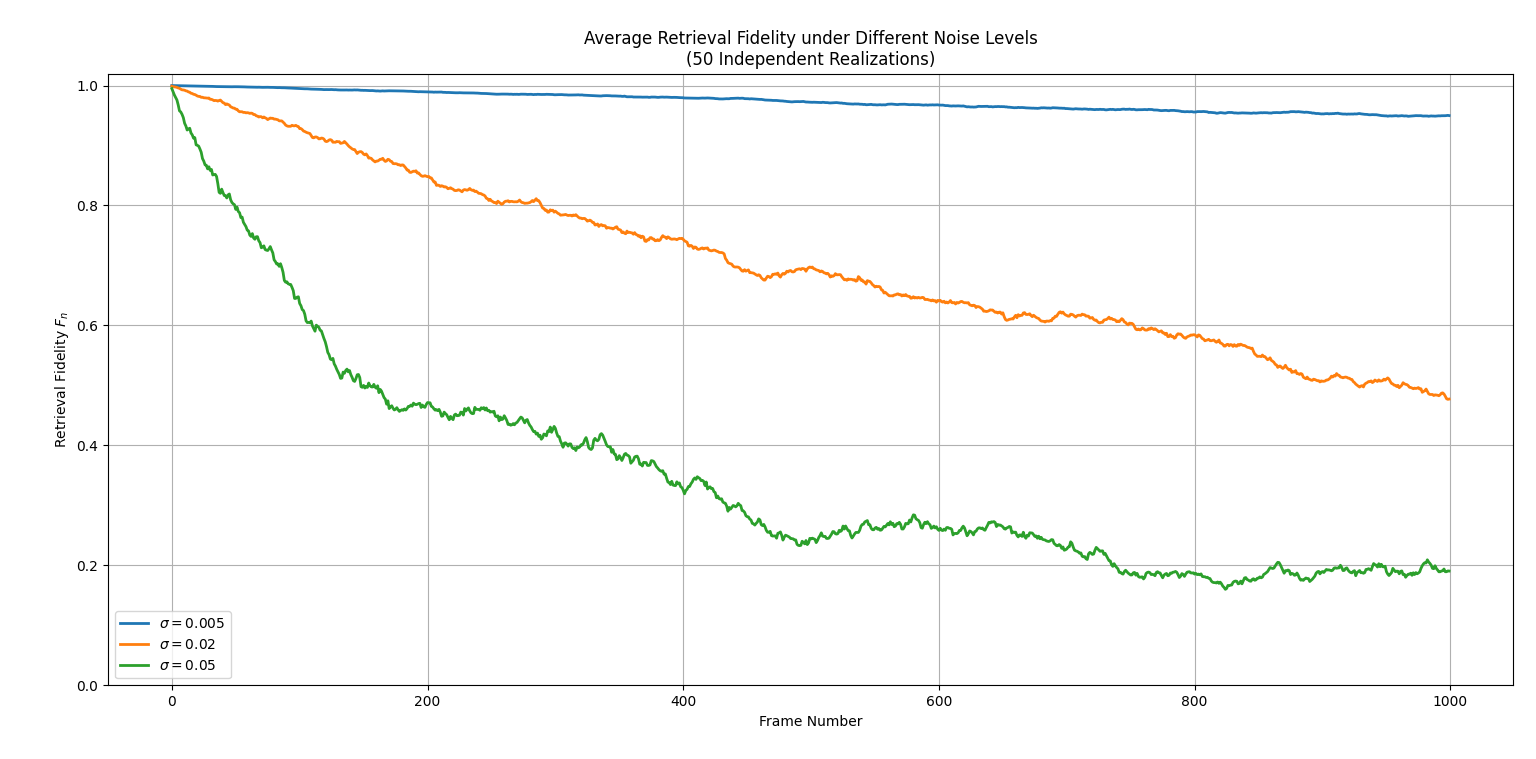}
\caption{
Average retrieval fidelity under different noise levels, averaged over 50 independent realizations. Higher noise amplitudes lead to faster degradation of the stored latent representation and reduced long-term retrieval accuracy.
}
\caption*{\footnotesize
Simulation parameters: $\eta=0.99$, $G=1/\eta$, 1000 sequential memory states, averaged over 50 independent realizations. Three physical noise levels were considered: $\sigma_{\epsilon}=0.005$, $0.020$, and $0.050$. Larger values of $\sigma_{\epsilon}$ correspond to stronger accumulated optical perturbations and therefore lower retrieval fidelity.
}
\label{fig:fidelity_noise}
\end{figure}

Figure~\ref{fig:fidelity_noise} shows the average retrieval fidelity for three different noise levels, namely $\sigma=0.005$, $\sigma=0.02$, and $\sigma=0.05$. In all the cases, the fidelity decreases with increasing storage duration due to the effect of optical loss, amplifier noise, and control imperfections. However, the rate of degradation strongly depends on the magnitude of the noise term.
\vspace{0.1cm}

For the lowest noise level ($\sigma=0.005$), the fidelity remains close to unity throughout the storage process, indicating that the encoded latent information is preserved with minimal distortion. Increasing the noise level to $\sigma=0.02$ results in a more pronounced fidelity decay, reflecting the gradual accumulation of stochastic errors. The highest noise level ($\sigma=0.05$) exhibits the fastest degradation, demonstrating the sensitivity of long-duration storage to uncontrolled noise sources.

\vspace{0.1cm}
The observed behavior is consistent with the evolution model, where noise accumulates over successive storage cycles and progressively increases the deviation between the ideal and stored coherent-state amplitudes. But, even in the presence of moderate noise, the memory retains a significant fraction of the encoded information over extended storage durations.
\vspace{0.1cm}

These results highlight the fundamental trade-off between storage duration and retrieval accuracy in continuous-variable photonic memories. 

\subsection{Metadata-Assisted Rollback Retrieval Performance}

The results of the previous subsection demonstrate that retrieval fidelity gradually decreases with storage duration, but the proposed architecture incorporates an additional rollback mechanism designed to recover historical memory states. Instead of storing independent copies of every frame, the system maintains a classical metadata database containing displacement histories, frame indices, entropy values, and auxiliary retrieval descriptors. Together, these records form a digital twin of the photonic memory evolution and enable reconstruction of previously stored latent states.
\vspace{0.1cm}

Assume that the memory currently stores the state corresponding to frame $N$. To retrieve a historical frame $m<N$, the controller accesses the displacement history associated with all subsequent updates and sequentially applies inverse displacement operations. The rollback operation can be expressed as

\begin{equation}
\alpha_m^{(\mathrm{rb})}
=
D(-\Delta\alpha_{m+1})
D(-\Delta\alpha_{m+2})
\cdots
D(-\Delta\alpha_N)
\alpha_N,
\end{equation}

where $\alpha_N$ denotes the current memory state and $D(-\Delta\alpha_k)$ represents the inverse displacement associated with the $k$-th storage update. By successively reversing the stored updates, the controller reconstructs an approximation of the latent state corresponding to frame $m$.
\vspace{0.1cm}

\begin{figure}[!t]
\centering
\includegraphics[width=\columnwidth]{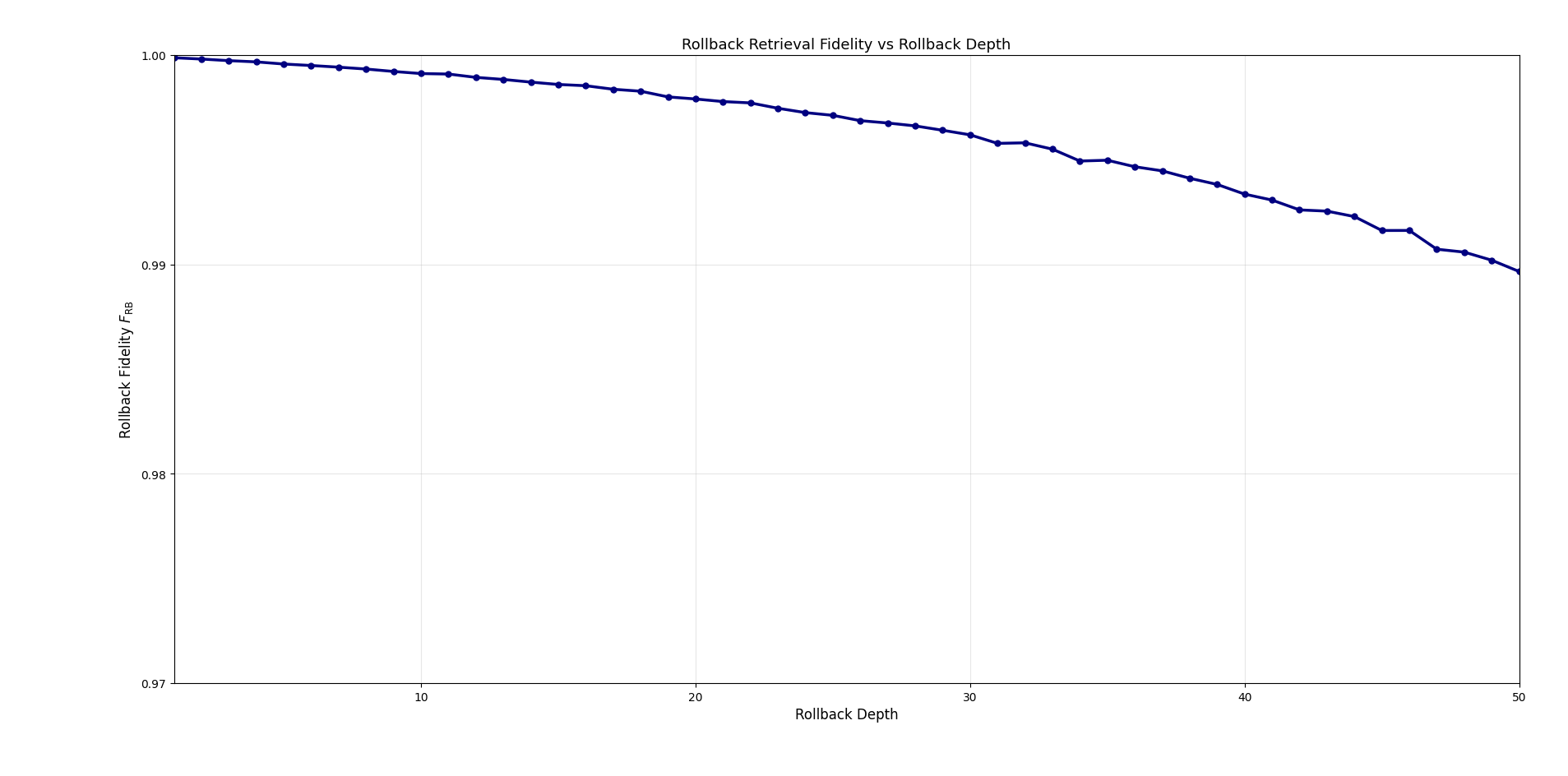}
\caption{
Average rollback retrieval fidelity as a function of rollback depth for the proposed eight-qumode photonic memory. The results are averaged over 500 independent realizations. Despite increasing rollback depth, the fidelity remains high, demonstrating the effectiveness of metadata-assisted state reconstruction.
}
\caption*{\footnotesize
Simulation parameters: $\lambda=0.98$, $\sigma_{\epsilon}=0.008$, $\sigma_{\mathrm{est}}=0.008$.The average rollback retrieval fidelity at the maximum rollback depth ($D=50$) is $F_{\mathrm{RB}}=0.9896$.
}
\label{fig:rollback_fidelity}
\end{figure}

Figure~\ref{fig:rollback_fidelity} presents the average rollback retrieval fidelity as a function of rollback depth. Here, the rollback depth corresponds to the number of displacement updates that must be reversed in order to reach the desired historical frame. The results were averaged over 500 independent realizations of the storage process.

\vspace{0.1cm}

The primary objective of the proposed rollback mechanism is to recover a previously stored coherent-state frame using metadata-assisted inverse displacement operations. Unlike conventional quantum memories that retrieve only the latest stored state, the proposed framework allows reversible access to arbitrary historical frames. Consequently, a quantitative measure of rollback accuracy is required.

Let the coherent-state evolution be governed by

\begin{equation}
\alpha_{n+1}
=
\lambda\alpha_n
+
\Delta_n
+
\epsilon_n,
\label{eq:forward}
\end{equation}

where

\begin{itemize}
\item $\lambda=\sqrt{\eta G}$ denotes the effective propagation coefficient determined by the loop transmissivity $\eta$ and optical amplifier gain $G$,
\item $\Delta_n$ represents the intentional displacement applied during the $n^{\mathrm{th}}$ update,
\item $\epsilon$ denotes the effective physical noise accumulated during the update.
\end{itemize}

The digital twin continuously monitors the photonic memory using sparse homodyne measurements. Rather than performing full quantum-state tomography, only coarse quadrature estimates are periodically recorded. Consequently, the digital twin stores an estimate

\begin{equation}
\hat{\alpha}_m
=
\alpha_m
+
\varepsilon_{\mathrm{est}},
\label{eq:digitaltwin}
\end{equation}

where $\alpha_m$ is the actual coherent-state amplitude corresponding to the desired rollback frame and $\varepsilon_{\mathrm{est}}$ denotes the homodyne estimation error.

The rollback process starts from the latest stored state, i.e.,

\begin{equation}
\alpha_n^{RB}=\alpha_n.
\end{equation}

Using Eq.~($47$), the first rollback step is

\begin{equation}
\alpha_{n-1}^{RB}
=
\frac{\alpha_n-\Delta_n}{\lambda}.
\end{equation}
\vspace{0.1cm}

Since $\epsilon_n$ represents stochastic physical perturbations that are neither known nor recorded during memory evolution, the rollback procedure reverses only the deterministic displacement history, while the accumulated perturbations remain in the reconstructed state.
\vspace{0.1cm}

Substituting the forward evolution from Eq.~($47$),

\begin{equation}
\alpha_n
=
\lambda\alpha_{n-1}
+
\Delta_n
+
\epsilon_n,
\end{equation}

gives

\begin{equation}
\begin{aligned}
\alpha_{n-1}^{RB}
&=
\frac{\lambda\alpha_{n-1}+\Delta_n+\epsilon_n-\Delta_n}{\lambda}\\
&=
\alpha_{n-1}
+
\frac{\epsilon_n}{\lambda}.
\end{aligned}
\end{equation}

Applying the rollback operation once more,

\begin{equation}
\alpha_{n-2}^{RB}
=
\frac{\alpha_{n-1}^{RB}-\Delta_{n-1}}{\lambda},
\end{equation}

and substituting

\begin{equation}
\alpha_{n-1}
=
\lambda\alpha_{n-2}
+
\Delta_{n-1}
+
\epsilon_{n-1},
\end{equation}

yields

\begin{equation}
\begin{aligned}
\alpha_{n-2}^{RB}
&=
\frac{
\lambda\alpha_{n-2}
+\Delta_{n-1}
+\epsilon_{n-1}
+\dfrac{\epsilon_n}{\lambda}
-\Delta_{n-1}
}{\lambda}\\
&=
\alpha_{n-2}
+
\frac{\epsilon_{n-1}}{\lambda}
+
\frac{\epsilon_n}{\lambda^2}.
\end{aligned}
\end{equation}

Repeating the above recursion until the desired rollback frame $m$ is reached produces

\begin{equation}
{
\alpha_{RB}
=
\alpha_m
+
\sum_{i=m+1}^{n}
\frac{\epsilon_i}{\lambda^{\,i-m}}.
}
\label{eq:rollbackstate}
\end{equation}

The rollback reconstruction error is therefore defined as the difference between the digital twin estimate and the reconstructed rollback state,

\begin{equation}
\delta
=
\hat{\alpha}_m
-
\alpha_{\mathrm{RB}}.
\label{eq:delta}
\end{equation}

Substituting Eqs.~(\ref{eq:digitaltwin}) and (\ref{eq:rollbackstate}) yields

\begin{equation}
{
\delta
=
\varepsilon_{\mathrm{est}}
-
\sum_{i=m+1}^{n}
\frac{\epsilon_i}{\lambda^{\,i-m}}.
}
\label{eq:deltafinal}
\end{equation}

Equation~(\ref{eq:deltafinal}) demonstrates that rollback accuracy depends on two independent sources of uncertainty:

\begin{enumerate}
\item Digital twin estimation uncertainty arising from sparse homodyne monitoring,
\item Accumulated physical noise introduced after the desired rollback frame.
\end{enumerate}

Since coherent-state overlap satisfies

\begin{equation}
F
=
\left|
\langle
\alpha
|
\beta
\rangle
\right|^2
=
\exp\!\left(-|\alpha-\beta|^2\right),
\end{equation}

the rollback retrieval fidelity is obtained as

\begin{equation}
{
F_{\mathrm{RB}}
=
\exp\!\left(
-
\left|
\varepsilon_{\mathrm{est}}
-
\sum_{i=m+1}^{n}
\frac{\epsilon_i}{\lambda^{\,i-m}}
\right|^2
\right).
}
\label{eq:FRB}
\end{equation}

Equation~(\ref{eq:FRB}) represents the theoretical rollback fidelity of the proposed photonic memory architecture. The expression reveals that retrieval fidelity is jointly determined by the effective propagation factor $\lambda$, accumulated physical disturbances $e_i$, rollback depth $(n-m)$, and the digital twin estimation error $\varepsilon_{\mathrm{est}}$.
\vspace{0.1cm}

The rollback fidelity also reveals an important stability condition for long-depth retrieval. Since the rollback error is given by

\begin{equation}
\delta
=
\epsilon_{\mathrm{est}}
-
\sum_{i=m+1}^{n}
\frac{\epsilon_i}
{\lambda^{\,i-m}},
\end{equation}

where $\lambda=\sqrt{\eta G}$ denotes the effective propagation factor, the corresponding error variance becomes

\begin{equation}
\mathrm{Var}(\delta)
=
\sigma_{\mathrm{est}}^{2}
+
\sigma_{\epsilon}^{2}
\sum_{k=1}^{D}
\frac{1}{\lambda^{2k}},
\end{equation}

with

\begin{equation}
D=n-m
\end{equation}

representing the rollback depth.
\vspace{0.15cm}

Equation $58$ shows that the accumulated rollback uncertainty increases monotonically with retrieval depth. In particular, when $\lambda<1$, successive inverse propagation amplifies previously accumulated perturbations, causing the rollback variance to grow rapidly for large values of $D$. Consequently, reliable historical retrieval requires the effective propagation factor to remain sufficiently close to unity so that rollback errors remain bounded.
\vspace{0.1cm}

This naturally defines a practical rollback horizon, beyond which the accumulated uncertainty causes the rollback fidelity to decrease below an acceptable operating threshold. Therefore, although rollback remains theoretically reversible, its achievable depth is ultimately limited by the residual propagation mismatch and the aggregate physical noise accumulated during memory evolution.
\vspace{0.1cm}

For numerical analysis, the effective propagation factor $\lambda$, physical update noise $\epsilon_i$, and digital twin estimation error $\varepsilon_{\mathrm{est}}$ are treated as tunable parameters. The resulting rollback fidelity is evaluated as a function of rollback depth, effective propagation coefficient, physical noise, and digital twin estimation accuracy.

%-------------------------------------------------
% Insert Simulation Figures Here
%-------------------------------------------------

% Figure 1 : Rollback Fidelity vs Rollback Depth

% Figure 2 : Rollback Fidelity vs Effective Propagation Factor

% Figure 3 : Rollback Fidelity vs Physical Noise

% Figure 4 : Rollback Fidelity vs Digital Twin Estimation Error

\section{Discussion}

\subsection{Memory Lifetime Analysis}

The retrieval fidelity curves presented in Fig.~\ref{fig:fidelity_noise} indicate that the performance of the proposed memory architecture gradually degrades as noise accumulates over storage cycles. A useful metric to describe long-term memory performance is the memory lifetime, defined as the maximum number of storage updates that can be performed before the retrieval fidelity falls below a specific threshold.
\vspace{0.1cm}

Assuming that stochastic perturbations accumulate approximately independently during the storage process, the mean-square deviation between the ideal and stored coherent state amplitudes can be approximated as

\begin{equation}
\left\langle
|\delta_n|^2
\right\rangle
\approx
n\sigma^2,
\end{equation}

where $\sigma^2$ denotes the effective noise variance per storage cycle and

\begin{equation}
\delta_n
=
\alpha_n^{\mathrm{ideal}}
-
\tilde{\alpha}_n
\end{equation}

represents the deviation between the ideal and noisy coherent-state amplitudes.
\vspace{0.1cm}

Using the coherent-state fidelity expression,

\begin{equation}
F_n
=
\exp
\left(
-
|\delta_n|^2
\right),
\end{equation}

the average retrieval fidelity may be approximated by

\begin{equation}
\langle F_n\rangle
\approx
\exp(-n\sigma^2).
\label{eq:lifetime_decay}
\end{equation}

In order to quantify the useful operating duration of the memory, a fidelity threshold $F_{\mathrm{th}}$ is introduced. The memory lifetime $N_{\mathrm{life}}$ is then defined as the largest storage depth for which the retrieval fidelity remains above this threshold,

\begin{equation}
N_{\mathrm{life}}
=
\max
\left\{
n :
F_n \geq F_{\mathrm{th}}
\right\}.
\end{equation}

Substituting Eq.~(\ref{eq:lifetime_decay}) yields

\begin{equation}
F_{\mathrm{th}}
=
\exp
\left(
-N_{\mathrm{life}}\sigma^2
\right),
\end{equation}

from which the memory lifetime can be expressed as

\begin{equation}
N_{\mathrm{life}}
=
-
\frac{\ln(F_{\mathrm{th}})}
{\sigma^2}.
\label{eq:memory_lifetime}
\end{equation}

This equation~(\ref{eq:memory_lifetime}) provides a simple analytical estimate of the storage capacity in terms of temporal depth. The result shows that the achievable memory lifetime increases rapidly as the effective noise level is reduced. So, improvements in optical transmission efficiency, amplifier performance, and control precision directly translate into longer storage durations and enhanced retrieval accuracy.
\vspace{0.1cm}

The analysis further suggests that the proposed architecture can support substantially longer memory depths when operated in low-noise regimes, consistent with the observation in the numerical fidelity simulations of Fig.~\ref{fig:fidelity_noise}.

\subsection{Rollback State Distinguishability}

While rollback retrieval fidelity quantifies the accuracy of reconstructing a historical memory state, an equally important characteristic is the number of historical states that can be uniquely distinguished during rollback retrieval.

\vspace{0.1cm}

To characterize this property, we model the rollback history as an undirected graph

\begin{equation}
G=(V,E),
\end{equation}

where each vertex $v_i\in V$ represents a stored rollback state corresponding to a coherent-state memory frame.
\vspace{0.1cm}

To quantify the distinguishability between two rollback states, we define the pairwise rollback-state fidelity as

\begin{equation}
F_{ij}
=
\left|
\left\langle
\alpha_{\mathrm{RB}}^{(i)}
\middle|
\alpha_{\mathrm{RB}}^{(j)}
\right\rangle
\right|^{2},
\end{equation}

where $\alpha_{\mathrm{RB}}^{(i)}$ and $\alpha_{\mathrm{RB}}^{(j)}$ denote the rollback coherent states corresponding to the $i^{\mathrm{th}}$ and $j^{\mathrm{th}}$ stored memory states, respectively. Since coherent states are not mutually orthogonal, rollback states with sufficiently large overlap become experimentally indistinguishable. Accordingly, an edge is introduced in the distinguishability graph whenever

\begin{equation}
(v_i,v_j)\in E
\iff
F_{ij}
\ge
F_{\mathrm{th}},
\end{equation}

where $F_{\mathrm{th}}$ denotes the minimum distinguishability threshold. The Rollback Distinguishability Capacity is then obtained from the maximum independent set of the resulting graph.

\vspace{0.1cm}

A valid rollback memory should consist only of mutually distinguishable rollback states. In graph-theoretic terms, this corresponds to an independent set of the rollback graph. Therefore, the maximum number of mutually distinguishable rollback states is given by the Maximum Independent Set (MIS),

\begin{equation}
D_{RB}
=
\alpha(G),
\end{equation}

where $\alpha(G)$ denotes the independence number of the rollback graph.

The Rollback Distinguishability Capacity is therefore defined as \cite{nielsen2010}

\begin{equation}
C_{RSD}
=
\log_{2}D_{RB}
=
\log_{2}\alpha(G),
\end{equation}

which represents the maximum rollback information that can be uniquely retrieved from the stored memory history.
\vspace{0.1cm}

Although determining the Maximum Independent Set is NP-hard for arbitrary graphs, the rollback graph of a finite photonic memory contains only a finite number of stored states. Consequently, the rollback distinguishability number can be obtained exactly for small memories through exhaustive search or by using standard Maximum Independent Set algorithms.
\vspace{0.1cm}

As a simple illustration, consider a $2\times2$ image consisting of four rollback states,

\begin{equation}
V=\{v_1,v_2,v_3,v_4\}.
\end{equation}

Suppose the rollback graph contains the edges

\begin{equation}
E=
\{
(v_1,v_2),
(v_2,v_3),
(v_3,v_4)
\},
\end{equation}
forming a simple path graph.
\vspace{0.1cm}

The Maximum Independent Set is

\begin{equation}
\{v_1,v_3\},
\end{equation}

or equivalently

\begin{equation}
\{v_2,v_4\},
\end{equation}

giving

\begin{equation}
D_{RB}=2.
\end{equation}

Hence, the Rollback Distinguishability Capacity becomes

\begin{equation}
C_{RSD}
=
\log_2(2)
=
1~\text{bit}.
\end{equation}

The value obtained in this illustrative example should not be interpreted as the storage capacity of the proposed photonic memory. Instead, $C_{\mathrm{RSD}}$ quantifies the distinguishability of historical rollback states represented by the graph. Specifically, a value of $C_{\mathrm{RSD}}=1$ bit indicates that the largest mutually distinguishable subset contains two rollback states, requiring one binary decision to uniquely identify a particular historical state. As the number of distinguishable rollback states increases, $\alpha(G)$ grows accordingly, leading to larger values of $C_{\mathrm{RSD}}$.
\vspace{0.1cm}

Thus, $C_{\mathrm{RSD}}$ serves as a graph-theoretic metric for quantifying the distinguishability of historical rollback states, providing an estimate of the maximum number of rollback states that can be uniquely resolved under a prescribed fidelity threshold.

\section{Conclusion}

This work has presented a comprehensive theoretical framework for reversible coherent-state photonic memory by integrating storage, retrieval, rollback dynamics, and digital-twin-assisted reconstruction within a unified analytical model. Beginning from the evolution of coherent states in an optical loop, we derived the complete mathematical description of state propagation under optical amplification, attenuation, and displacement operations. The resulting framework not only quantifies the fidelity of direct retrieval but also establishes a physically consistent rollback mechanism capable of recovering previously stored memory states through successive inverse displacement operations. Unlike conventional photonic memories, which are primarily designed for sequential storage and readout, the proposed architecture introduces temporal accessibility as an intrinsic property of the memory itself, thereby enabling controlled access to historical information without compromising the integrity of the most recently stored state.
\vspace{0.1cm}

A key contribution of this work is the incorporation of a digital twin that separates logical reconstruction from physical rollback. Since the digital twin continuously tracks the evolution of the coherent-state amplitudes using experimentally acquired metadata, complete image or video reconstruction can be performed entirely within the digital domain while leaving the physical optical memory undisturbed. Physical rollback is therefore required only when the historical optical state itself is of interest, rather than the reconstructed information. This distinction significantly improves the practicality of the proposed architecture by reducing unnecessary physical operations and minimizing additional degradation arising from repeated rollback procedures. The analytical expressions developed for retrieval fidelity and rollback retrieval fidelity further provide quantitative measures for evaluating the robustness of both the storage process and the rollback mechanism under realistic noisy operating conditions.
\vspace{0.1cm}

Beyond fidelity analysis, this work also establishes an information-theoretic characterization of reversible photonic memory through rollback state distinguishability (RSD).\cite{cerf2007} The proposed formulation provides a quantitative estimate of the maximum number of mutually distinguishable rollback states that can coexist within a given fidelity threshold. This graph-theoretic interpretation extends conventional storage-capacity analysis by incorporating retrieval reliability directly into the capacity measure, thereby linking physical noise, rollback fidelity, and information accessibility within a common mathematical framework. Consequently, the proposed capacity formulation serves not merely as a measure of stored information but also as an indicator of the practical recoverability of historical memory states.
\vspace{0.1cm}

The present study nevertheless represents a theoretical foundation rather than a complete experimental realization. Several important directions remain open for future investigation. In particular, experimentally accurate quantum-noise models for optical parametric amplification, including amplifier-added vacuum fluctuations and phase instability, should be incorporated to obtain tighter bounds on rollback performance under realistic operating conditions. Likewise, a rigorous stability analysis relating rollback depth, propagation factor, and accumulated noise variance would establish fundamental operational limits for reversible photonic memories. Future work will also investigate hardware implementation using integrated photonic circuits, programmable optical delay loops, and high-speed electro-optic modulation to experimentally validate the proposed framework.
\vspace{0.1cm}

While the present architecture has been developed for coherent-state memories, its mathematical formulation naturally provides a pathway toward genuinely quantum photonic memories based on non-classical continuous-variable states. Extending the rollback formalism to squeezed states, multimode Gaussian resource states, and entangled photonic systems may enable capabilities beyond those accessible through classical digital twins, potentially introducing genuine quantum advantages \cite{kok2007} for reversible information storage and retrieval.

%%%%%%%%%%%%%%%%%%%%%%%%%%%%%%%%%%%%%%%%%%%%%%%%%%%%%%%%%%%%%%%%%%%%%%%%%%%%%%%%

%%%%%%%%%%%%%%%%%%%%%%%%%%%%%%%%%%%%%%%%%%%%%%%%%%%%%%%%%%%%%%%%%%%%%%%%%%%%%%%%

%%%%%%%%%%%%%%%%%%%%%%%%%%%%%%%%%%%%%%%%%%%%%%%%%%%%%%%%%%%%%%%%%%%%%%%%%%%%%%%%
\section*{Acknowledgment}

The author would like to thank the Department of Physics, Sardar Vallabhbhai National Institute of Technology (SVNIT), Surat, for providing an academic environment that encouraged independent research and scientific exploration. The author is also grateful to the faculty members whose courses and discussions contributed to the development of the ideas presented in this work.
\vspace{0.1cm}

Appreciation also extends to the availability of open scientific literature, computational resources, and modern research tools that facilitated the theoretical analysis and numerical simulations reported in this paper.

\section*{Declarations}

\subsection*{Funding}

The author received no financial support from any funding agency in the public, commercial, or not-for-profit sectors for the research, authorship, or publication of this work.

\subsection*{Competing Interests}

The author declares that there are no competing financial or non-financial interests related to this work.

\subsection*{Availability of Data and Materials}

This study is based on theoretical analysis and numerical simulations. No experimental datasets were generated or analysed. The numerical simulation code and data generated during this study are available from the corresponding author upon reasonable request for research and verification purposes.

\subsection*{Author Contributions}

The author conceived the research, developed the theoretical framework, performed the mathematical derivations and numerical simulations, analyzed the results, and wrote the manuscript.

\bibliographystyle{IEEEtran}
\bibliography{Bibliography/PaperBibliography}

@book{nielsen2010,
  author = {Michael A. Nielsen and Isaac L. Chuang},
  title = {Quantum Computation and Quantum Information},
  publisher = {Cambridge University Press},
  edition = {10th Anniversary Edition},
  year = {2010}
}

@book{gerry2005,
  author = {Christopher C. Gerry and Peter L. Knight},
  title = {Introductory Quantum Optics},
  publisher = {Cambridge University Press},
  year = {2005}
}

@book{ghatak1998,
  author = {Ajoy Ghatak and K. Thyagarajan},
  title = {Introduction to Fiber Optics},
  publisher = {Cambridge University Press},
  year = {1998}
}

@misc{optec,
  author = {{Op-Tec}},
  title = {Integrated Photonics},
  howpublished = {\url{https://op-tec.org}},
  year = {2025}
}

@misc{sanjit2025,
  author = {Sanjit Krishna and S. Krishna},
  title = {Qumode-Based Quantum Image Storage Using Continuous-Variable Photonic Encoding},
  year = {2025},
  eprint = {2507.03290},
  archivePrefix = {arXiv},
  primaryClass = {quant-ph}
}

@article{weedbrook2012,
  author = {Christian Weedbrook and Stefano Pirandola and Ra\'ul Garc\'ia-Patr\'on and Nicolas J. Cerf and Timothy C. Ralph and Jeffrey H. Shapiro and Seth Lloyd},
  title = {Gaussian Quantum Information},
  journal = {Reviews of Modern Physics},
  volume = {84},
  pages = {621--669},
  year = {2012}
}

@article{braunstein2005,
  author = {Samuel L. Braunstein and Peter van Loock},
  title = {Quantum Information with Continuous Variables},
  journal = {Reviews of Modern Physics},
  volume = {77},
  pages = {513--577},
  year = {2005}
}

@article{caves1982,
  author = {Carlton M. Caves},
  title = {Quantum Limits on Noise in Linear Amplifiers},
  journal = {Physical Review D},
  volume = {26},
  number = {8},
  pages = {1817--1839},
  year = {1982}
}

@article{silverstone2016,
  author = {Joshua W. Silverstone and Domenico Bonneau and Mark G. Thompson and Jeremy L. O'Brien},
  title = {Silicon Quantum Photonics},
  journal = {IEEE Journal of Selected Topics in Quantum Electronics},
  volume = {22},
  number = {6},
  pages = {390--402},
  year = {2016}
}

@article{cerf2007,
  author = {Nicolas J. Cerf and Gerd Leuchs and Eugene S. Polzik},
  title = {Quantum Information with Continuous Variables of Atoms and Light},
  journal = {Imperial College Press},
  year = {2007}
}

@article{kok2007,
  author = {Pieter Kok and W. J. Munro and Kae Nemoto and Timothy C. Ralph and Jonathan P. Dowling and Gerard J. Milburn},
  title = {Linear Optical Quantum Computing with Photonic Qubits},
  journal = {Reviews of Modern Physics},
  volume = {79},
  pages = {135--174},
  year = {2007}
}

\end{document}